\documentclass[pra,aps,showpacs,superscriptaddress,twocolumn]{revtex4-2}
\usepackage{graphicx,amsmath,amssymb,color}
\usepackage{xcolor}
\usepackage{tabularx}
\usepackage[colorlinks, citecolor={blue!50!black}, urlcolor={blue!50!black}, linkcolor={red!50!black}]{hyperref}
\begin{document}

\title{Entangled atomic ensemble and an yttrium-iron-garnet sphere in coupled microwave cavities}

\author{Dilawaiz}
\affiliation{Department of Physics and Applied Mathematics, Pakistan Institute of Engineering and Applied Sciences (PIEAS), Nilore $45650$, Islamabad, Pakistan.}
\author{Shahid Qamar}
\affiliation{Department of Physics and Applied Mathematics, Pakistan Institute of Engineering and Applied Sciences (PIEAS), Nilore $45650$, Islamabad, Pakistan.}
\affiliation{Center for Mathematical Sciences, PIEAS, Nilore, Islamabad $45650$, Pakistan.}
\author{Muhammad Irfan}
\email[Corresponding author: ]{m.irfanphy@gmail.com}
\affiliation{Department of Physics and Applied Mathematics, Pakistan Institute of Engineering and Applied Sciences (PIEAS), Nilore $45650$, Islamabad, Pakistan.}
\affiliation{Center for Mathematical Sciences, PIEAS, Nilore, Islamabad $45650$, Pakistan.}

\date{\today}

\begin{abstract}
We present a scheme to generate distant bipartite and tripartite entanglement between an atomic ensemble and a yttrium iron garnet (YIG) sphere in coupled microwave cavities.
We consider atomic ensemble in a single-mode microwave cavity which is coupled with a second single-mode cavity having a YIG sphere.
Our system, therefore, has five excitation modes namely cavity-1 photons, atomic ensemble, cavity-2 photons, a magnon and a phonon mode in the YIG sphere.
We show that significant bipartite entanglement exists between indirectly coupled subsystems in the cavities, which is robust against temperature.
Moreover, we present suitable parameters for a significant tripartite entanglement of ensemble, magnon, and phonon modes.
We also demonstrate the existence of tripartite entanglement between magnon and phonon modes of the YIG sphere with indirectly coupled cavity photons.
Interestingly, this distant tripartite entanglement is of the same order as previously found for a single-cavity system.
We show that cavity-cavity coupling strength affects both the degree and transfer of quantum entanglement between various subsystems.
Therefore, an appropriate cavity-cavity coupling optimizes the distant entanglement by increasing the entanglement strength and its robustness against temperature.
\end{abstract}
\maketitle
\newpage
\section{Introduction}
Quantum entanglement is recognized as the most fascinating aspect of quantum formalism~\cite{horodecki_quantum_2009}.
It has applications in quantum information processing, quantum networking, quantum dense coding, quantum-enhanced metrology, and so on~\cite{frowis_macroscopic_2018, kimble_quantum_2008,simon_towards_2017,moller_quantum_2017}.
Therefore, its realization through physical resources used in information processing and communication protocols necessitates a scale above the subatomic level for the ease of experimental implementation~\cite{lukin_quantum_2001}.
That is why there is growing attention toward the exploration of quantum mechanical effects at the macroscopic level.
The advancement in micro and nanofabrication in recent years provided novel platforms to study macroscopic entanglement.
Cavity optomechanics is one such system that received a lot of attention during the past decade~\cite{aspelmeyerRevModPhys, favero_focus_2014}.
Among other applications~\cite{favero_optomechanics_2009, meystre_short_2013}, cavity optomechanics enables quantum state transfer between different modes of electromagnetic fields~\cite{wang_using_2012,mari_opto-_2012} which has a central role in quantum information processing networks. 
Moreover, a possible platform for quantum information processing is offered by atomic ensembles.
They can serve as valuable memory nodes for quantum communication networks due to their longer coherence duration and collective amplification effect~\cite{sangouard_quantum_2011}.  
Another promising physical platform is yttrium iron garnet (YIG), a ferrimagnetic material, due to its high spin density and low decay rates of collective spin excitations (i.e., Kittel mode~\cite{tabuchi_hybridizing_2014}), resulting in the strong coupling between Kittel mode and cavity photons~\cite{huebl_high_2013,wang_magnetization_2020,goryachev_high-cooperativity_2014,bai_spin_2015}.  
\par
Since the initial experiments, many hybrid quantum systems based on quantum magnonics have been studied for their possible applications in quantum technologies~\cite{hisatomi_bidirectional_2016,osada_cavity_2016,haigh_triple-resonant_2016,lachance-quirion_resolving_2017}.
Magnon Cavity QED is a relatively newer field and a potential candidate for studying new features of strong-coupling QED.
The observation of bi-stability and the single superconducting qubit coupling to the Kittel mode are interesting developments in this field~\cite{wang_bistability_2018,tabuchi_coherent_2015}. Li \textit{et al.} illustrated how to create tripartite entanglement in a system of microwave cavity photons entangled to the magnon and phonon modes of a YIG sphere in a magnomechanical cavity~\cite{li_magnon-photon-phonon_2018}.
This study was followed by an investigation of magnon-magnon entanglement between two YIG spheres in cavity magnomechanics~\cite{li_entangling_2019}.
Later, Wu \textit{et al.} investigated magnon-magnon entanglement between two YIG spheres in cavity optomagnonics~\cite{wu_remote_2021}.
Likewise, Ning and Yin theoretically demonstrated the entanglement of magnon and superconducting qubit utilizing a two-mode squeezed-vacuum microwave field in coupled cavities~\cite{ning_entangling_2021}.
Wang \textit{et al.} explored nonreciprocal transmission and entanglement in two-cavity magnomechanical system~\cite{wang_nonreciprocal_2022}.
That work was succeeded by studying a long-range generation of magnon-magnon entangled states via qubits~\cite{ren_long-range_2022}.
\par
Potential schemes for distant entanglement between disparate systems are increasingly considered for testing fundamental limits to quantum theory and possible applications in quantum networks~\cite{chen_macroscopic_2013}.
In an interesting study, Joshi \textit{et al.} theoretically examined whether two spatially distant cavities connected by an optical fiber may produce quantum entanglement between mechanical and optical modes~\cite{joshi_entanglement_2012}.
Likewise, many researchers theoretically explored other schemes for transferring entanglement at a distance which includes an array of three optomechanical cavities for the study of the entanglement between different mechanical and optical modes~\cite{akram_photon-phonon_2012} and a doubly resonant cavity with a gain medium of cascading three-level atoms placed in it to investigate entanglement transfer from two-mode fields to the two movable mirrors~\cite{ge_entanglement_2013}. 
In a double cavity optomechanical system, Liao \textit{et al.} quantified the entanglement of macroscopic mechanical resonators by the concurrence~\cite{liao_entangling_2014}.
It was followed by a study of entanglement transfer from the inter-cavity photon-photon entanglement  to an intracavity photon-phonon via two macroscopic mechanical resonators~\cite{rehaily_entanglement_2017}.
Recently, Bai \textit{et al.} proposed a scheme of a two-cavity coupled optomechanical system with the atomic ensemble and a movable mirror in distinct cavities, through which they showed ensemble-mirror entanglement and entanglement transfer between different subsystems~\cite{bai_robust_2016}. 
\par
In the past, several cavity optomechanical systems have been studied for entanglement with atomic medium~\cite{li_manipulating_2020, bai_robust_2016,ian_cavity_2008,zhou_entanglement_2011}.
Recently, it is shown that atomic ensemble can be entangled with magnon modes within a single cavity~\cite{Fan-2023, wu-2023}.
However, to the best of our knowledge, distant entanglement of atomic ensemble and YIG sphere in microwave cavities has not been reported yet.
In this paper, we present a method for entangling atomic ensemble to mechanical and Kittel modes in YIG sphere placed within coupled microwave cavities.
In our study, we have considered an atomic ensemble containing $N\sim10^{7}$~\cite{genes_emergence_2008,hald_spin_1999} atoms and YIG sphere with a typical diameter of $250\,{\mu}$m ~\cite{zhang_cavity_2016}, a promising platform to study distant macroscopic entanglement.
We show that significant bipartite and tripartite entanglement exists between the magnon and phonon modes of YIG sphere placed in cavity-2 with the atomic ensemble and cavity-1 photons.
It is interesting to find that YIG sphere can be entangled to an indirectly coupled cavity field.
We illustrate that this distant entanglement can be controlled by varying cavity-cavity coupling strength. 
Since atomic ensembles can serve as efficient memory nodes for quantum communication networks, we therefor believe that the considered hybrid system has useful applications in quantum technologies.
\section{system model and Hamiltonian}
\begin{figure}[htb] 
\includegraphics[width=1.\linewidth]{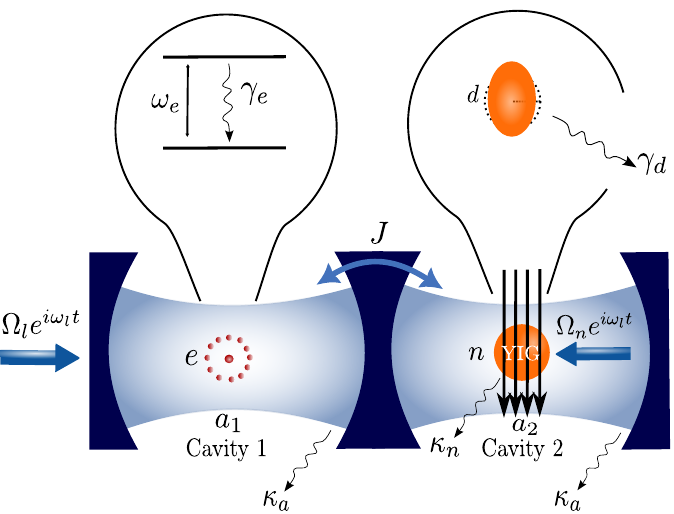}
\caption{\label{fig:1}
Schematic representation of two single-mode cavities coupled to each other with coupling strength $J$ incorporating an atomic ensemble of $N$ two-level atoms characterized by intrinsic frequency $\omega_e$ placed in the cavity 1 and a YIG sphere placed in cavity 2. An external laser field drives the cavity at frequency $\omega_l$ with strength $\Omega_l$. Correspondingly, a microwave magnetic field at frequency $\omega_l$ with strength $\Omega_n$ drives the magnon modes of the YIG sphere, enhancing the magnomechanical coupling. YIG sphere is concurrently influenced by the cavity's magnetic field, bias magnetic field,  and the drive magnetic field, all orthogonal to each other at the site of the YIG sphere. The decay rates of cavity modes ($a_1$ and $a_2$), atomic ensemble ($e$), magnon mode ($n$), and phonon mode ($d$) associated with the YIG sphere are given by $\kappa_a$ ,$\gamma_e$, $\kappa_n$, and $\gamma_d$, respectively.}
\end{figure}
We consider a hybrid coupled-cavity magnomechanical system which consists of two single-mode cavities with resonance frequency $\omega_k$ $(k=1,2)$ encasing an atomic ensemble and a YIG sphere as shown in Fig.~\ref{fig:1}.
This coupled system has five excitation modes namely microwave electromagnetic modes in cavity 1 and cavity 2, magnon and phonon modes in the YIG sphere, and atomic excitation in cavity 1.

In cavity 2, a YIG sphere is placed close to the maximum magnetic field of the cavity mode and is simultaneously acted upon by a bias magnetic field, thus establishing the photon-magnon coupling.
The external bias magnetic field excites the magnon modes. 
The magnetic field of the cavity mode interacts with Kittle mode via magnetic dipole interaction, in which spins evenly precess in the ferrimagnetic sphere.
The bias field \textbf{B} and the gyromagnetic ratio $\Gamma$ control the magnon frequency, i.e., $\omega_{n}=\Gamma \textbf{B}$.
Varying magnetization in the YIG sphere results in magnetostriction leading to the interplay of energy between magnon and phonon modes in it. 

In cavity 1, an ensemble of $N$ two-level atoms with transition frequency $\omega_e$ interacts with the cavity field. 
The atoms constituting the ensemble are individually characterized by the spin-1/2 Pauli matrices $\sigma_{+}, \sigma_{-}$, and $\sigma_{z}$. 
Collective spin operators of atomic polarization for the atomic ensemble are described as $S_{+,-, z}=\sum_{i=1}^{N} \sigma_{+,-, z}^{(i)}$, and they follow the commutation relations $\left[S_{+}, S_{-}\right]=S_{z}$ and $\left[S_{z}, S_{\pm}\right]=\pm 2 S_{\pm}$~\cite{bai_robust_2016}.
The operators $S_{\pm}$ and $S_{z}$ may be represented in terms of the bosonic annihilation and creation operators $e$ and $e^{\dagger}$ by using the Holstein-Primakoff transformation~\cite{zheng_generation_2012,holstein_field_1940,hammerer_quantum_2010}: $S_{+}=e^{\dagger} \sqrt{N-e^{\dagger} e} \simeq \sqrt{N} e^{\dagger}$,
$S_{-}=\sqrt{N-e^{\dagger} e} e \simeq \sqrt{N} e$, $S_{z}=e^{\dagger} e-N / 2$, where $e$ and $e^{\dagger}$ follow the commutation relation $\left[e, e^{\dagger}\right]=1$.
This transformation is valid only when the population of atoms in the ground state is large compared to the atoms in the excited state, so that $S_z \simeq\left\langle S_z\right\rangle \simeq-N$~\cite{genes_emergence_2008}.

To simplify our analysis, we have considered both the frequency of the drive laser field and the drive magnetic field to be $\omega_l$.
The Hamiltonian describing the system under rotating-wave approximation in a frame rotating with the frequency of the drive fields ($\omega_l$) is given by:
\begin{align}\label{E1}
 H/\hbar=& \sum_{k=1}^{2}  \Delta_{k} a_{k}^{\dagger} a_{k}+ \Delta_{e} e^{\dagger}e + \Delta_{n} n^{\dagger} n+ \frac{ \omega_{d}}{2}\left(x^{2}+y^{2}\right) \nonumber \\
&+ g_{n d} n^{\dagger} nx + J\left(a_{1}^{\dagger} a_{2}+a_{1} a_{2}^{\dagger}\right)+G_{a e}\left(e a_{1}^{\dagger}+e^{\dagger} a_{1}\right) \nonumber \\
 & +g_{n a}\left(a_{2} n^{\dagger}+a_{2}^{\dagger} n\right)+i  \Omega_{l} \left(a_{1}^{\dagger} -a_{1} \right) \nonumber \\
 & +i  \Omega_{n} \left(n^{\dagger} -n \right).
\end{align}
\noindent
In Eq.~(\ref{E1}), the energy associated with the cavity (1 and 2), atomic excitation, and magnon modes is represented in the first three terms where $a_k\left(a_k^{\dagger}\right)$, $e\left(e^{\dagger}\right)$, and $n\left(n^{\dagger}\right)$ are the annihilation (creation) operators of the cavity, collective atomic excitation, and magnon mode, respectively.
Here, $\Delta_{k}$ ($k=1,2$), $\Delta_{e}$, and $\Delta_{n}$ are the detunings of cavity mode's frequency $\omega_k$ ($k=1,2$), the intrinsic frequency of two-level atoms in the atomic ensemble $\omega_{e}$, and magnon mode's frequency $\omega_n$ with respect to the drive field's frequency $\omega_l$, i.e., $\Delta_{k}=\omega_{k}-\omega_{l}$ ($k=1,2$), $\Delta_{e}=\omega_{e}-\omega_{l}$, and $\Delta_{n}=\omega_{n}-\omega_{l}$.
Rotating-wave approximation holds when $\omega_{k} (k=1,2), \omega_{n}, \omega_{e} \gg$ $g_{n a}, \kappa_{a}, \kappa_{n}, \gamma_{e}$ (which is satisfied in cavity magnomechanics by experimentally feasible parameters)~\cite{zhang_strongly_2014}.
The fourth term in Eq.~(\ref{E1}) is the energy of the mechanical mode (phonon mode) of frequency $\omega_d$ with dimensionless position $x$ and momentum $y$ operators satisfying $[x, y]=i$.
The next four terms describe the interaction of all coupled subsystems in the cavity system encompassing coupling of magnon and phonon, cavity 1 and cavity 2, collective atomic excitation and cavity 1, cavity 2 and magnon modes with strength $g_{n d}$, $J$, $G_{a e}$, and $g_{n a}$, respectively.
The magnomechanical coupling strength $g_{n d}$, resulting from magnetostrictive interaction, is typically weak, however, it can be enhanced by the drive microwave field having frequency $\omega_l$ applied at the site of YIG.
The coupling rate of collective atomic excitation with cavity mode $G_{a e}=g \sqrt{N}$ where $g$ is the atom-cavity coupling strength defined as $g=\nu \sqrt{\omega_{1} / 2 \hbar \epsilon_{0} V}$, with $\nu$ the dipole moment of atomic transition, $V$ the volume of the cavity, and $\epsilon_{0}$ the permittivity of free space.
Regarding the eight-term (magnon-photon coupling), when the coherent energy exchange rate between light and matter is faster than their decay rates, the coupling strength between magnon and photon reaches the strong coupling regime, i.e., $g_{n a}>\kappa_a,\kappa_n$~\cite{huebl_high_2013,tabuchi_hybridizing_2014,zhang_strongly_2014,goryachev_high-cooperativity_2014,bai_spin_2015}.
The second-last term describes a microwave field driving cavity 1 with Rabi frequency $\Omega_l=\sqrt{2 {P\kappa_a} / \hbar \omega_l}$ which depends on the input power $P$ of the drive field and the decay rate $\kappa_a$ of the cavity.
Similarly, we also consider a magnon-mode drive field (last term in Eq.~(\ref{E1})) with Rabi frequency $\Omega_n=\frac{\sqrt{5}}{4} \Gamma \sqrt{N_s} B_0$ with $\Gamma$ the gyromagnetic ratio, $N_s$ the total number of spins, and $B_0$ the applied field's amplitude.
In case of YIG sphere,  $\Gamma/2\pi=28$~GHz/T, and $N_s=\rho V_s$ with volume $V_s$ and spin density $\rho=4.22 \times 10^{27} \mathrm{~m}^{-3}$ of the sphere~\cite{li_magnon-photon-phonon_2018}.
We assume low-lying excitations while deriving $\Omega_n$, i.e., $\left\langle n^{\dagger} n\right\rangle \ll 2 N_s \varsigma$, where $\varsigma=\frac{5}{2}$ is the spin number of the ground state $\mathrm{Fe}^{3+}$ ion in YIG~\cite{li_magnon-photon-phonon_2018}.
At a temperature $T$, the equilibrium mean thermal photon [$a_k$ ($k=1,2$)], magnon ($n$), and phonon ($d$) number is given by $ Z_{h}(\omega_{h})=[\exp (\hbar \omega_{h} / k_{B} T)-1]^{-1}$ $(h=a_k,n,d)$, where $k_{B}$ is the Boltzmann constant.

Since, we are interested to study steady-state quantum entanglement in the linear regime we therefore, use standard input-output theory resulting in quantum Langevin equations (QLEs) where the effect of the input noise operator is added for each excitation mode:
%
\begin{align}\label{E2}
&\dot{a}_1=-\left(\kappa_a+i \Delta_1\right) a_1-i G_{a e} e-i J a_2+\sqrt{2 \kappa_a} a_1^{i n}+\Omega_{l}, \nonumber \\
&\dot{a}_2=-\left(\kappa_a+i \Delta_2 \right) a_2 -i J a_1-i g_{n a} n+\sqrt{2 \kappa_a} a_2^{i n}, \nonumber \\
&\dot{e}=-\left(\gamma_{e}+i \Delta_{e}\right) e-i G_{a e} a_1+\sqrt{2 \gamma_{e}} e^{i n},\nonumber \nonumber\\
&\dot{n}=-\left(i \Delta_{n}+\kappa_{n}\right) n-i g_{n a} a_2-i g_{n d} n x+\Omega_{n}+\sqrt{2 \kappa_{n}} n^{i n},\nonumber \\
&\dot{x}=\omega_{d} y,\;
\dot{y}=-\omega_{d} x-\gamma_{d} y-g_{n d} n^{\dagger} n+\xi,\end{align}
\noindent
with zero-mean input noise operators $a_k^{in}$, $e^{in}$, $n^{in}$ and $\xi$  for $k^{th}$ cavity, atomic-excitation, magnon, and phonon modes, respectively.
The parameters $\gamma_e$ and $\gamma_d$ are the atomic decay rate and the mechanical damping rate, respectively.
The input noise operators under Markovian approximation, which is valid for large mechanical quality factor, $Q=\omega_{d} / \gamma_{d} \gg 1$ are characterized by the following non-vanishing correlation functions that are $\delta-$correlated in time domain~\cite{clerk_introduction_2010}: $\langle a_{1}^{\text {in }}(\tau) a_{1}^{i n \dagger}(\tau^{\prime})\rangle=[Z_{a_{1}}(\omega_{a_{1}})+1] \delta(\tau-\tau^{\prime})$, $\langle a_{1}^{i n \dagger}(\tau) a_{1}^{\text {in }}(\tau^{\prime})\rangle=Z_{a_{1}}(\omega_{a_{1}}) \delta(\tau-\tau^{\prime})$, $\langle a_{2}^{\text {in }}(\tau) a_{2}^{i n \dagger}(\tau^{\prime}t)\rangle=[Z_{a_{2}}(\omega_{a_{2}})+1] \delta(\tau-\tau^{\prime})$, 
$\langle a_{2}^{i n \dagger}(\tau) a_{2}^{\text {in }}(\tau^{\prime})\rangle=Z_{a_{2}}(\omega_{a_{2}}) \delta(\tau-\tau^{\prime})$, $\langle e^{i n}(\tau) e^{i n \dagger}(\tau^{\prime})\rangle=\delta(\tau-\tau^{\prime})$, $\langle n^{\text {in }}(\tau) n^{\text {in } \dagger}(\tau^{\prime})\rangle=[Z_{n}(\omega_{n})+1] \delta(\tau-\tau^{\prime})$, $\langle n^{\text {in } \dagger}(\tau) n^{i n}(\tau^{\prime})\rangle=Z_{n}(\omega_{n})\delta(\tau-\tau^{\prime})$,  and $\langle\xi(\tau) \xi(\tau^{\prime})+\xi(\tau^{\prime})\xi(\tau)\rangle / 2 \simeq \gamma_{d}[2 Z_{d}(\omega_{d})+1] \delta(\tau-\tau^{\prime})$, where $\tau$ and $\tau^\prime$ denote two distinct times.
It is important to note that the $\delta-$correlated mechanical noise approximation is only valid for a large mechanical quality factor.
For the case of low-quality factor, we have to solve for the exact correlation function of the noise operators.
\begin{widetext}
\onecolumngrid

From the quantum Langevin Eq.~\eqref{E2}, we obtain the expressions for the steady-state values of cavities, ensemble, magnon, and phonon mode operators given by:
\begin{equation}
\begin{aligned}
&\left\langle a_{1}\right\rangle=\frac{\Omega_{l}\left(\kappa_a + i \Delta_2 \right)\left(\kappa_n+i\tilde{\Delta}_n\right)\left(\gamma_e + i \Delta_e \right)+ g_{n a}^2\Omega_{l} \left(\gamma_e + i \Delta_e \right) -g_{n a}  \Omega_n J\left(\gamma_e + i \Delta_e \right)  } {S}, \\
&\left\langle a_{2}\right\rangle=\frac{-i J \left(\kappa_n+i\tilde{\Delta}_n\right)\left\langle a_{1}\right\rangle-i g_{n a} \Omega_n }{\left(\kappa_a + i \Delta_2 \right)\left(\kappa_n+i\tilde{\Delta}_n\right)+ g_{n a}^2},\,\,\langle e\rangle=\frac{-i G_{a e}\left\langle a_{1}\right\rangle}{\left(\gamma_{e}+i \Delta_{e}\right)}, \,\,\langle n\rangle=\frac{\Omega_{n}-i g_{n a}\left\langle a_{2}\right\rangle}{i \tilde{\Delta}_{n}+\kappa_{n}}, \\
&\langle x\rangle=-\left(\frac{g_{n d}}{\omega_{d}}\right)|\langle n\rangle|^{2}, \,\,\langle y\rangle=0,
\end{aligned}
\end{equation}
where
\\
$S=(\kappa_{a}+i \Delta_{1})(\gamma_{e}+$ $i \Delta_{e})[(\kappa_{a}+i \Delta_{2})
(\kappa_{n}+$ $i \tilde{\Delta}_{n}
)+$ $g_{n a}^{2}]$ $-G_{a e}^2[(\kappa_{a}+i \Delta_{2})$ $(\kappa_{n}+i \tilde{\Delta}_{n})$ $+g_{n a}^{2}]$ $+J^{2}(\gamma_{e}+i \Delta_{e})(\kappa_{n}$ $+i \tilde{\Delta}_{n})$, \\
and the effective magnon detuning $\Tilde{\Delta}_{n}=\Delta_{n}+g_{n d}\langle x\rangle$.
The effective magnomechanical coupling rate is $G_{nd}=i\sqrt{2}g_{nd}\langle n\rangle$.
\end{widetext}
\twocolumngrid
To analyze the steady-state entanglement of the system, we linearize the dynamics of the coupled cavity system.
We assume that the cavity is intensely driven with a very high input power, resulting in significant steady-state amplitudes for the intracavity fields and magnon modes, respectively, i.e., $\left|\langle a_{k}\rangle\right| \gg 1$ ($k=1,2$)~\cite{genes_robust_2008} and $|\langle n\rangle| \gg 1$~\cite{li_magnon-photon-phonon_2018}.
For a proper choice of drive field's reference phase, $\langle a_k \rangle$ may be treated real~\cite{genes_robust_2008}.
Moreover, the bosonic description of atomic polarization may only be used when the single-atom excitation probability is noticeably below 1.
The conditions of large steady-state amplitudes of intracavity fields and low excitation limit of atoms in the ensemble are simultaneously satisfied only when $g^{2} /\left(\Delta_{e}^{2}+\gamma_{e}^{2}\right) \ll\left|\langle a_1 \rangle\right|^{-2} \ll 1$.
This necessitates a weak atom-cavity coupling~\cite{genes_emergence_2008}.
Hence, in the strong driving limit, we can neglect the second-order fluctuation terms, such that the operator $P$ ($P=a_k,e,n,x,y$) can be written as $P=\langle P\rangle+\delta P$ where $\langle P\rangle$ represents the steady-state part while $\delta P$ represents the zero-mean fluctuation associated with $P$.
In the opposite limit where the quantum effects of a single or few excitations are important~\cite{Nunnenkamp-2021, Ludwig_2008} or in studying fully nonlinear Hamiltonian~\cite{Qvarfort_PRA}, the standard quantum master equation may be used to study the dynamics of the system.
Similarly, a coupling with a non-equilibrium environment also requires exact quantum Langevin equations~\cite{LudwigPhysRevA}.
Next, we define quadrature fluctuations ($\delta U_1(t)$, $\delta W_1(t)$, $\delta U_2(t)$, $\delta W_2(t)$, $\delta u_1(t)$, $\delta w_1(t)$, $\delta x(t)$, $\delta y(t)$, $\delta u_2(t)$, $\delta w_2(t)$), with $\delta U_{1}=\left(\delta a_1+\delta a_1^{\dagger}\right) / \sqrt{2}$, $\delta W_{1}=i\left(\delta a_1^{\dagger}-\delta a_1\right) / \sqrt{2}$,
$\delta U_{2}=(\delta a_2+\delta a_2^{\dagger}) / \sqrt{2}$,
$\delta W_{2}=i(\delta a_2^{\dagger}-\delta a_2) / \sqrt{2}$,
$\delta u_{1}=(\delta n+\delta n^{\dagger}) / \sqrt{2}$,
$\delta w_{1}=i(\delta n^{\dagger}-\delta n) / \sqrt{2}$,
$\delta u_{2}=(\delta e+\delta e^{\dagger}) / \sqrt{2}\,$ and
$\delta w_{2}=i(\delta e^{\dagger}-\delta e) / \sqrt{2}$, 
to work out a set of linearized quantum Langevin equations 
\begin{align}\label{E3}
\dot{r}(t)=A r(t)+o(t);
\end{align}
with $r(t)$ the fluctuation operator in the form of quadrature fluctuations: $r(t)^T=[\delta U_1(t), \delta W_1(t), \delta U_2(t),$ $\delta W_2(t), \delta u_1(t),$ $\delta w_1(t), \delta x(t),$ $\delta y(t),\delta u_2(t),$ $\delta w_2(t)]$, $o(t)$ denotes the noise operators represented as: $o(t)^T=[\sqrt{2 \kappa_{a}}$ $U_{1}^{i n}(t),$ $\sqrt{2 \kappa_{a}}$ $W_{1}^{i n}(t),$ $\sqrt{2 \kappa_{a}} U_{2}^{i n}(t),$ $\sqrt{2 \kappa_{a}} W_{2}^{i n}(t),$ $\sqrt{2 \kappa_{n}} u_{1}^{i n}(t),$ $\sqrt{2 \kappa_{n}} w_{1}^{i n}(t),$ $0$, $\xi(t),
\,\sqrt{2 \gamma_{e}}u_{2}^{i n}(t),$ $\sqrt{2 \gamma_{e}} w_{2}^{i n}(t)]$, and $A$ is the drift matrix:

\begin{widetext}
\onecolumngrid
\begin{align}\label{E4}
A=\begin{pmatrix}
-\kappa_{a} & \Delta_{1} & 0 & J & 0 & 0 & 0 & 0 & 0 & G_{a e} \\ -\Delta_{1} & -\kappa_{a} & -J & 0 & 0 & 0 & 0 & 0 & -G_{a e} & 0 \\ 0 & J & -\kappa_{a} & \Delta_{2} & 0 & g_{n a} & 0 & 0 & 0 & 0 \\ -J & 0 & -\Delta_{2} & -\kappa_{a} & -g_{n a} & 0 & 0 & 0 & 0 & 0 \\ 0 & 0 & 0 & g_{n a} & -\kappa_{n} & \tilde{\Delta}_{n} & -G_{n d} & 0 & 0 & 0 \\ 0 & 0 & -g_{n a} & 0 & -\tilde{\Delta}_{n} & -\kappa_{n} & 0 & 0 & 0 & 0 \\ 0 & 0 & 0 & 0 & 0 & 0 & 0 & \omega_{d} & 0 & 0 \\ 0 & 0 & 0 & 0 & 0 & G_{nd} & -\omega_{d} & -\gamma_{d} & 0 & 0 \\ 0 & G_{a e} & 0 & 0 & 0 & 0 & 0 & 0 & -\gamma_{e} & \Delta_{e} \\ -G_{a e} & 0 & 0 & 0 & 0 & 0 & 0 & 0 & -\Delta_{e} & -\gamma_{e}
\end{pmatrix}.
\end{align}
\end{widetext}
\twocolumngrid
\noindent
The linearized quantum Langevin equations [see Eq.~(\ref{E3})] correspond to an effective linearized Hamiltonian, which ensures the Gaussian state of the system when it is stable.
Thus, the linearized dynamics of the system along with the Gaussian nature of the noises lead to the continuous-variable five-mode Gaussian state of the steady-states corresponding to its quantum fluctuations.
Routh-Hurwitz criterion is used to work out the stability conditions for our linearized system~\cite{dejesus_routh-hurwitz_1987}.
The system becomes stable and attains its steady-state only when real parts of all eigenvalues of the drift matrix ($A$) are negative.
The steady-state Covariance Matrix (CM), which describes the variance within each subsystem and the covariance across several subsystems, is generated from the following Lyapunov equation when the stability requirements are met~\cite{parks_1993}: 
 \begin{equation}
A\mathcal{V}+\mathcal{V} A^{T}=-D,
\end{equation}
where $D= \operatorname{diag}\,[\kappa_{a}(2 \,Z_{a}+1), \kappa_{a}(2 \,Z_{a}+1), \kappa_{a}(2\, Z_{a}+1),$
$\kappa_{a}(2\, Z_{a}+1)$, $\kappa_{n}\left(2 Z_{n}+1\right)$, $\kappa_{n}\left(2 Z_{n}+1\right)$, $0$, $\gamma_{d}(2\, Z_{d}+1), \gamma_{e}, \gamma_{e}]^{T}$ is the diffusion matrix, for the corresponding decays originating from the noise correlations.
\noindent
To quantify bipartite entanglement among different subsystems of the coupled two-cavity system, we use logarithmic negativity ($\mathit{E_{N}}$)~\cite{plenio_logarithmic_2005,vidal_computable_2002}. We have a five-mode Gaussian state characterized by a covariance matrix $\mathcal{V}$ which can be expressed in the form of a block matrix:
\begin{equation}
\mathcal{V}=\left(\begin{array}{ccccc}
\mathcal{V}_{a_1}  & \mathcal{V}_{a_1 a_2} & \mathcal{V}_{a_1 n} & \mathcal{V}_{a_1 d} & \mathcal{V}_{a_1 e}\\ 
\mathcal{V}^T_{a_1 a_2}  & \mathcal{V}_{a_2} & \mathcal{V}_{a_2 n} & \mathcal{V}_{a_2 d} & \mathcal{V}_{a_2 e}\\ 
{\mathcal{V}}^T_{a_1 n}  & \mathcal{V}^T_{a_2 n} & \mathcal{V}_{n} & \mathcal{V}_{n d} & \mathcal{V}_{n e}\\
\mathcal{V}^T_{a_1 d}  & \mathcal{V}^T_{a_2 d} & \mathcal{V}^T_{n d} & \mathcal{V}_{d} &\mathcal{V}_{d e}\\ 
\mathcal{V}^T_{a_1 e}  & \mathcal{V}^T_{a_2 e} & \mathcal{V}^T_{n e} & \mathcal{V}^T_{d e} & \mathcal{V}_{e}\\ 
\end{array}\right),
\label{covariance2}
\end{equation}
\noindent
where each block is a $2 \times 2$ matrix.
Here, diagonal blocks represent the variance within each subsystem [(cavity 1) photon, (cavity 2) photon, magnon, phonon, and ensemble].
The correlations between any two distinct degrees of freedom of the entire magnomechanical system are represented by the off-diagonal blocks, which are covariances across distinct subsystems~\cite{bai_robust_2016}. 
Following Simon’s criterion~\cite{simon_peres-horodecki_2000} to judge the non-separability of the transposed modes in the transposed submatrix derived from the covariance matrix $\mathcal{V}$, we compute logarithmic negativity numerically. 
The covariance matrix ($10 \times 10$) $\mathcal{V}$ is reduced to a submatrix $\mathcal{V}_{l}$ ($4 \times 4$) in order to evaluate the covariance between the subsystems.
For instance, the submatrix representing the covariance of cavity 1 and cavity 2 subsystems is determined by the first four rows and columns of $\mathcal{V}$.
We can represent $\mathcal{V}_{l}$ of cavity 1-cavity 2 subsystems in the following way~\cite{bai_robust_2016}:
\begin{equation}
\mathcal{V}_{l}=\left(\begin{array}{cc}
\mathcal{V}_{a_1} & \mathcal{V}_{a_1 a_2} \\
\mathcal{V}_{a_1 a_2}^{T} & \mathcal{V}_{a_2}
\end{array}\right),
\end{equation}
where $a_1$ index the cavity-1 subsystem and $a_2$ index the  cavity-2 subsystem.
Similarly, the covariance of other subsystems can be determined by considering their corresponding rows and columns in $\mathcal{V}$.
Then, transposed covariance sub-matrix $\tilde{\mathcal{V}_{l}}$ is obtained by partial transposition of $\mathcal{V}_{l}$ employing $\tilde{\mathcal{V}}_{l}=\mathcal{T}_{1 \mid 2} \mathcal{V}_{l} \mathcal{T}_{1 \mid 2}$, where $\mathcal{T}_{1 \mid 2}=\operatorname{diag}(1,-1,1,1)$ realizes partial transposition at the level of covariance matrices~\cite{simon_peres-horodecki_2000}.
Then, we compute the minimum symplectic eigenvalue $\tilde{f}_{-}$ of the transposed CM $\tilde{\mathcal{V}}_{l}$ using $\tilde{f}_{-}=\min \operatorname{eig}\left|i \Theta_{2} \tilde{\mathcal{V}}_{l}\right|$ with $\Theta_{2}=\oplus_{j=1}^{2} i \sigma_{y}$ and $\sigma_{y}$ the $y$-Pauli matrix~\cite{li_magnon-photon-phonon_2018}.
If the smallest eigenvalue is less than 1/2, the inseparability of the transposed modes is ensured, i.e., the modes are entangled. $\mathit{E_{N}}$ is evaluated as~\cite{vidal_computable_2002}:
 \begin{equation}
E_{N} \equiv \max \left[0,-\ln 2 \tilde{f}_{-}\right].
\label{ln}
\end{equation}
Similarly, residual contangle $\mathcal{R}_{\tau}^{min}$~\cite{adesso_continuous_2006}, which is a continuous variable analog of the tangle for discrete variable tripartite entanglement~\cite{coffman_distributed_2000}, is used for the quantification of tripartite entanglement, which is defined as~\cite{adesso_continuous_2006}:
\begin{equation}
\mathcal{R}_{\tau}^{\min } \equiv \min \left[\mathcal{R}_{\tau}^{o \mid n d}, \mathcal{R}_{\tau}^{n \mid o d}, \mathcal{R}_{\tau}^{d \mid o n}\right],
\label{residue}
\end{equation}
where $n$ stands for magnon whereas $d$ stands for phonon mode, $o=a_1$ for cavity-magnon-phonon tripartite entanglement, and $o=e$ for magnon-phonon-ensemble tripartite entanglement.
In Eq. (\ref{residue}) $\mathcal{R}_{\tau}^{k \mid l m}$ is evaluated using $\mathcal{R}_{\tau}^{k \mid l m} \equiv C_{k \mid l m}-C_{k \mid l}-C_{k \mid m}\,(k, l, m=o, n, d)$, $C_{k \mid l m}$ is the squared one-mode-vs-two-modes logarithmic negativity $\mathit{E}_{k \mid l m}$ and  $C_{k \mid l}$ is the contangle of subsystems of $k$ and $l$~\cite{li_magnon-photon-phonon_2018}, defined as the squared logarithmic negativity $\mathit{E}_{k \mid l}$~\cite{vidal_computable_2002}.
To compute $\mathit{E}_{k \mid l m}$ following the definition of logarithmic negativity given in Eq.(\ref{ln}), $\Theta_{2}=\oplus_{j=1}^{2} i \sigma_{y}$ is replaced by $\Theta_{3}=\oplus_{j=1}^{3} i \sigma_{y}$ and the transposed covariance matrix  $\tilde{\mathcal{V}}$ is obtained by carrying out the partial transposition of covariance matrix $\mathcal{V}$, i.e., $\tilde{\mathcal{V}}=\mathcal{T}_{k \mid l m} \mathcal{V} \mathcal{T}_{k \mid l m}$, where the partial transposition matrices ~\cite{li_magnon-photon-phonon_2018} are:
 $\mathcal{T}_{1 \mid 23}=\operatorname{diag}(1,-1,1,1,1,1)$, $\mathcal{T}_{2 \mid 13}=\operatorname{diag}(1,1,1,-1,1,1)$,
and  $\mathcal{T}_{3 \mid 12}=\operatorname{diag}(1,1,1,1,1,-1)$.
 
\section{Results and Discussion}
In this section, we present the results of our numerical simulations.
We have adopted the following experimentally feasible parameters for the system involving microwave cavities and YIG sphere in our simulations~\cite{li_magnon-photon-phonon_2018}: $\omega_k/{2\pi}=\omega_n/{2\pi}=10\,$GHz $(k=1,2)$, $\omega_d/{2\pi}=10\,$MHz, $\gamma_d/{2\pi}=10^2\,$Hz, $\kappa_a/{2\pi}=\kappa_n/{2\pi}=1\,$MHz, $g_{n a}/{2\pi}=3.2\,$MHz, $G_{n d}/{2\pi}=4.8\,$MHz, and temperature $T=10\,$mK.
Correspondingly, the atom-cavity coupling and atomic decay rate are considered of the order of megahertz, i.e, $G_{a e}/{2\pi}=6\,$MHz and $\gamma_e/{2\pi}=1\,$MHz.
Further, the hopping rate $J$ between the cavities is also of the order of megahertz. 
It can be seen that for the above-chosen parameters, our system is well within the low-excitation regime of atomic ensemble, satisfying the condition: $g^{2} /\left(\Delta_{e}^{2}+\gamma_{e}^{2}\right) \ll 1$.
\begin{table}[h!]
\begin{tabular}{|l|c|}
\hline
\textbf{Bipartite subsystems} & \textbf{Symbol for entanglement}\\

\hline
Cavity 1-magnon & $E_N^{a_1n}$\\
\hline
Cavity 1-phonon & $E_N^{a_1d}$\\
\hline

Cavity 2-magnon & $E_N^{a_2n}$\\
\hline
Cavity 2-phonon & $E_N^{a_2d}$\\
\hline
Magnon-ensemble & $E_N^{ne}$\\
\hline
Phonon-ensemble & $E_N^{de}$\\
\hline
Magnon-phonon & $E_N^{nd}$\\
\hline
\end{tabular}
\caption{Notation adopted for the representation of bipartite entanglement.}
\label{table:1}
\end{table}
\par
First, we discuss the results of bipartite entanglement.
We have five different modes in the coupled-cavity system; therefore, entanglement can exist in any combination of two modes. Interestingly, we observe promising results for macroscopic distant entanglement, i.e., the entanglement of atomic ensemble and cavity-1 photons with phonon and magnon modes of the YIG sphere placed in cavity 2. We also illustrate entanglement transfer from phonon-ensemble ($de$) and magnon-ensemble ($ne$) subsystems to cavity 1 photon-phonon ($a_1d$) and cavity 1 photon-magnon subsystems ($a_1n$) when detuning parameters and cavity-cavity coupling strength are changed.
In Table~\ref{table:1}, we have summarized the symbols we adopted in our simulations to represent the bipartite entanglement of different combinations of subsystems.
\begin{figure}[htb] 
	\includegraphics[width=\linewidth]{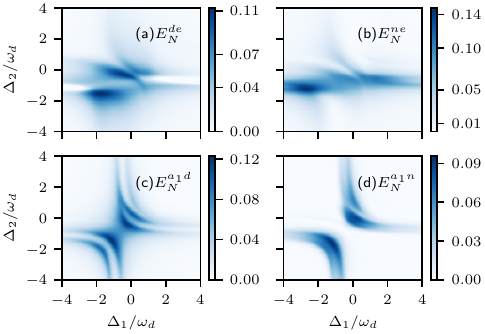}
	\caption{Density plot of bipartite entanglement (a) $E_N^{d e}$, (b) $E_N^{n e}$, (c) $E_N^{a_1d}$, and (d) $E_N^{a_1n}$ versus normalized cavity-1 detuning $\Delta_1/{\omega_d}$ and cavity-2 detuning $\Delta_2/{\omega_d}$. In (a)-(b) $\Delta_e=-\omega_d$ whereas in (c)-(d) $\Delta_e=\omega_d$. In all cases, $\tilde{\Delta}_n=0.9\,\omega_d$ and $J=0.8\,\omega_d$.}\label{fig:2}
\end{figure}

In Fig.~\ref{fig:2}, we present four different distant bipartite entanglements as a function of dimensionless detuning of the cavity 1 $\left(\Delta_1/\omega_d\right)$ and cavity 2 $\left(\Delta_2/\omega_d\right)$.
We have considered magnon detuning $\tilde{\Delta}_n$ to be $0.9\,\omega_d$ (near-resonant with blue sideband) while coupling between the two cavities is $J=0.8\,\omega_d$.
Fig.~\ref{fig:2}(a)-(b) illustrates ensemble-phonon ($E_N^{de}$) and ensemble-magnon ($E_N^{ne}$) entanglement for ensemble detuning $\Delta_e$ to be $-\omega_d$ (resonant with red sideband).
Although the ensemble and YIG sphere are placed in separate cavities, we find strong entanglement for both $E_N^{de}$ and $E_N^{ne}$. $E_N^{de}$ attains maximum value around $\Delta_2\approx-1.5\,\omega_d$ and $\Delta_2\approx0$ corresponding to $\Delta_1\approx-2\,\omega_d$ and $\Delta_1\approx-0.5\,\omega_d$. It can be seen that $E_N^{ne}$ is manifested primarily around $\Delta_2\,\approx-\omega_d$ in the entire range of $\Delta_1/\omega_d$. However, maximum $E_N^{ne}$ exists around $\Delta_1\approx-2.5\,\omega_d$. Similarly, we present cavity-1 photon-phonon ($E_N^{a_1d}$) and cavity-1 photon-magnon ($E_N^{a_1n}$) entanglement in Fig.~\ref{fig:2}(c)-(d) for $\Delta_e=\omega_d$. Both systems exhibit strong entanglement around $\Delta_1\approx-\omega_d$ and $\Delta_1\approx0$. If we follow $\Delta_1=\Delta_2$ line on the plane formed by $\Delta_1$ and $\Delta_2$, we observe that there are two distinct detuning regions for maximal $E_N$ on the density plots showing $E_N^{a_1d}$ and $E_N^{a_1n}$ compared to a single joint region along the $\Delta_1=-\Delta_2$ line. 
\par
For further analysis, we consider two cases.
In the first case, cavity 1 and cavity 2 have the same detuning frequency with respect to the frequency of the drive field, i.e., $\Delta_1=\Delta_2=\Delta_a$, (symmetric detuning).
If the first cavity is red-detuned or blue-detuned, the second cavity is also red-detuned or blue-detuned.
In the second case, cavity 1 and cavity 2 have opposite detuning frequencies with respect to the frequency of the drive field, i.e., $\Delta_1=-\Delta_2=-\Delta_a$, (non-symmetric detuning).
If the first cavity is red-detuned, the second is blue-detuned, and vice versa.
\begin{figure}[htb] 
	\includegraphics[width=\linewidth]{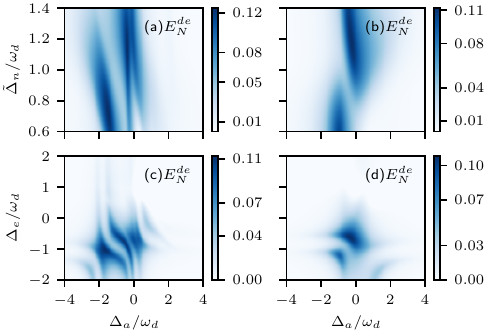}
	\caption{Density plot of $E_N^{d e}$ versus normalized cavity detuning $\Delta_a/{\omega_d}$ and (a)-(b) magnon detuning $\tilde{\Delta}_n/ {\omega_d}$ at $\Delta_e=-\omega_d$, and (c)-(d) ensemble detuning $\Delta_e/{\omega_d}$ at $\tilde{\Delta}_n=0.9\,\omega_d$. In (a) and (c) $\Delta_a/{\omega_d}=\Delta_1/{\omega_d}=\Delta_2/{\omega_d}$. However, $\Delta_a/{\omega_d}=-\Delta_1/{\omega_d}=\Delta_2/{\omega_d}$ in (c) and (d). The cavity-cavity coupling strength is taken to be $J=\omega_d$.}\label{fig:3}
\end{figure}

Next, we present phonon-ensemble entanglement ($E_N^{de}$) as a function of normalized cavity detuning $(\Delta_a/{\omega_d})$ against dimensionless magnon detuning $\tilde{\Delta}_n/ {\omega_d}$ in Fig.~\ref{fig:3}(a)-(b) and ensemble detuning $\Delta_e/{\omega_d}$ in Fig.~\ref{fig:3}(c)-(d).
In the left panel, we have symmetric cavity field detuning while in the right panel, the detuning is non-symmetric. It can be seen in Fig.~\ref{fig:3} (a)-(b) that significant entanglement is present for the complete range of effective magnon detuning.
We consider $0.6\leq\tilde{\Delta}_n/\omega_d\leq1.4$, where we get stronger entanglement. While $E_N^{de}$ is significant for the broad range of $\tilde{\Delta}_{n}$, it strongly depends on the choice of cavity field detuning.
There are two distinct regions of cavity detuning where we find maximum entanglement. One region is around cavity resonance for both the symmetric and non-symmetric choices of detuning, while the other region depends on the choice.
For the symmetric case, strong entanglement is also present around $\Delta_a\approx-1.75\,\omega_d$.
However, for the non-symmetric case, this second region is around $\Delta_a\approx-\omega_d$.
The lower panel in Fig.~\ref{fig:3} shows that $E_N^{de}$ is maximum around  $\Delta_e\approx-\omega_d$, while the choices of cavity detuning are approximately the same as discussed above in the previous case.

\begin{figure}[htb] 
	\includegraphics[width=\linewidth]{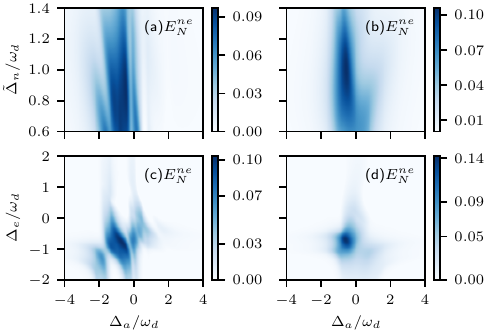}
	\caption{Density plot of $E_N^{n e}$ versus normalized cavity detuning $\Delta_a/\omega_d$ and (a)-(b) magnon detuning $\tilde{\Delta}_n/ {\omega_d}$ at $\Delta_e=-\omega_d$, and (c)-(d) ensemble detuning $\Delta_e/{\omega_d}$ at $\tilde{\Delta}_n=0.9\,\omega_d$. In (a) and (c) $\Delta_a/{\omega_d}=\Delta_1/{\omega_d}=\Delta_2/{\omega_d}$. However, $\Delta_a/{\omega_d}=-\Delta_1/{\omega_d}=\Delta_2/{\omega_d}$ in (b) and (d). The cavity-cavity coupling strength is taken to be $J=0.8\,\omega_d$.}\label{fig:4}
\end{figure}

Fig.~\ref{fig:4} shows magnon-ensemble entanglement ($E_N^{ne}$) as a function of normalized cavity detuning against dimensionless magnon detuning (upper panel) and ensemble detuning (lower panel).
$E_N^{ne}$ is optimal around $\Delta_a\approx-0.5\,\omega_d$. Fig.~\ref{fig:4}(a)-(b) shows that $E_N^{ne}$ is significant for the whole range of $\tilde{\Delta}_{n}$ while it is maximum around $\Delta_e\approx-\omega_d$ as shown in Fig.~\ref{fig:4}(c)-(d).
In both cases, we note that entanglement exists for a wider parameter space in symmetric detuning as compared to the non-symmetric detuning choice.
Similar to $E_N^{de}$,  $E_N^{ne}$ is also significant around $\Delta_e\approx-\omega_d$ and $\tilde{\Delta}_n\approx0.9\,\omega_d$.
As a result, we conclude that the bipartite entanglement of modes involving the atomic ensemble and YIG sphere is most remarkable when the magnon is near-resonant with the anti-Stokes band while the ensemble is resonant with the Stokes band.

\begin{figure}[t]
	\includegraphics[width=\linewidth]{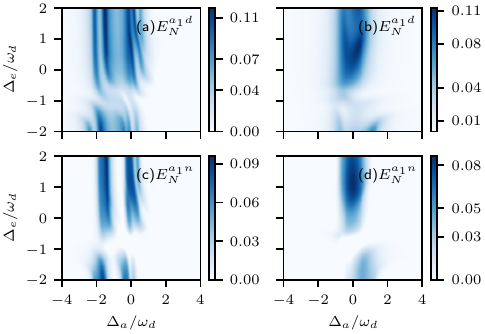}
	\caption{Density plot of $E_N^{a_1 d}$ and $E_N^{a_1 n}$ versus normalized cavity detuning $\Delta_a/\omega_d$ and ensemble detuning $\Delta_e/{\omega_d}$ at $J=0.8\,\omega_d$ and $\tilde{\Delta}_n=0.9\,\omega_d$. In (a) and (c) $\Delta_a/{\omega_d}=\Delta_1/{\omega_d}=\Delta_2/{\omega_d}$. However, $\Delta_a/{\omega_d}=-\Delta_1/{\omega_d}=\Delta_2/{\omega_d}$ in (b) and (d).}\label{fig:5}
\end{figure}

Fig.~\ref{fig:5} shows cavity-1 photon-phonon entanglement $E_N^{a_1d}$ and cavity-1 photon-magnon entanglement $E_N^{a_1n}$ as a function of normalized cavity detuning and ensemble detuning.
The left panel is for symmetric detuning whereas the right panel is for non-symmetric detuning.
For the symmetric case, we have significant $E_N^{a_1d}$ and $E_N^{a_1n}$ around $\Delta_a\approx-2\,\omega_d$ and $\Delta_a\approx 0$ for a wide range of $\Delta_e$ [see Fig.~\ref{fig:5}(a) and (c)].
For the second case [see Fig.~\ref{fig:5}(b) and (d)], $E_N^{a_1d}$ and $E_N^{a_1n}$ are significant around resonance frequency of both cavities.
In contrast to $E_N^{de}$ and $E_N^{ne}$, both $E_N^{a_1d}$ and $E_N^{a_1n}$ are prominent when the atomic ensemble is resonant with the anti-Stokes sideband, i.e, at $\Delta_e\approx\omega_d$ and almost negligible at $\Delta_e=-\omega_d$ (the Stokes sideband) as depicted in Fig.~\ref{fig:5}. 
\par
\begin{figure}[htb]
	\includegraphics[width=\linewidth]{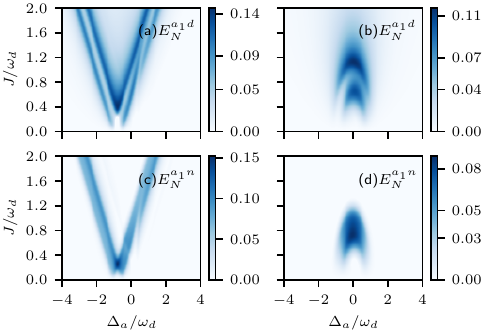}
	\caption{Density plot of $E_N^{a_1 d}$ and $E_N^{a_1 n}$ versus normalized cavity detuning $\Delta_a/\omega_d$ and coupling strength $J/\omega_d$ at $\tilde{\Delta}_n=0.9\,\omega_d$ and $\Delta_e=\omega_d$. In (a) and (c) $\Delta_a/{\omega_d}=\Delta_1/{\omega_d}=\Delta_2/{\omega_d}$. However, $\Delta_a/{\omega_d}=-\Delta_1/{\omega_d}=\Delta_2/{\omega_d}$ in (b) and (d).}\label{fig:6}
\end{figure}

We illustrate the dependence of cavity-1 photon-phonon entanglement ($E_N^{a_1d}$) and cavity-1 photon-magnon entanglement ($E_N^{a_1n}$) on cavity-cavity coupling strength $J$ and cavity detuning $\Delta_a$ in Fig.~\ref{fig:6}, where we choose $\tilde{\Delta}_n=0.9\,\omega_d$ and $\Delta_e=\omega_d$.
As expected, the bipartite entanglement of these subsystems is non-existent in the absence of cavity-cavity coupling.
For the symmetric cavity detuning [see Fig.~\ref{fig:6}(a) and (c)], entanglement first increases with increasing $J$ around $\Delta_a\approx-0.5\,\omega_d$; however, beyond a certain value, any further increase in $J$ shifts the detuning region for optimal entanglement to the right and left of $\Delta_a\approx-0.5\,\omega_d$.
However, for the non-symmetric detuning [see Fig.~\ref{fig:6}(b) and (d)] the trend is quite different.
Here, $E_N^{a_1d}$ and $E_N^{a_1n}$ increase with increasing coupling strength $J$ till a particular value.
We note that the dependence on $J$ varies when different values of $\Delta_e$ and $\tilde{\Delta}_n$ are considered.
For the given parameters, $E_N^{a_1d}$ first increases as a function of $J$ reaching a local maximum at $J\approx0.65\,\omega_d$ at resonance followed by a downtrend from $J\approx0.65\,\omega_d$ to $J\approx 0.9\omega_d$, after which it increases again up till $J\approx1.12\,\omega_d$ and decreases afterwards.
On the other hand, $E_N^{a_1n}$ attains maximum value from $J\approx0.5\,\omega_d$  to $J\approx0.75\,\omega_d$, then it decreases gradually up till $J\approx1.25\,\omega_d$ before dying out thereafter.
It is important to note that there is a downtrend in $E_N^{a_1d}$ around $J=0.9\,\omega_d$, which gives a significant value for $E_N^{a_1n}$.
The reason lies in the entanglement transfer between the different subsystems, which is further elaborated in the following analysis.

To study entanglement transfer, we set $-\Delta_1=\Delta_2=\Delta_a$ in our simulations.
The role of $J$ in the degree and dynamics of entanglement transfer between different subsystems is further elaborated in Fig.~\ref{fig:7}.
At smaller values of cavity-cavity coupling $J$, the cavity-2 modes are significantly entangled (See $E_N^{nd}$, $E_N^{a_2d}$ and $E_N^{a_2n}$) around $\Delta_2=-\omega_d$.
When the coupling is increased, the cavity-1 photon and cavity-2 photon interact with each other, resulting in a redistribution of cavity photon excitations which translates to the other excitation modes.
For instance, at $\Delta_2=-\omega_d$, $E_N^{nd}$ and $E_N^{a_2d}$ decrease with increasing cavity-cavity coupling while most of the other bipartite entanglements increase [See Appendix~\ref{A} for details].
This transfer not only decreases with increasing $J$ but there is also a corresponding decrease in the strength of $E_N^{a_2d}$ and $E_N^{a_2n}$.
This decrease accounts for the corresponding increase in $E_N^{de}$, $E_N^{ne}$, and $E_N^{a_1d}$.
Another interesting feature is that at smaller $J$, maximum entanglement of $E_N^{a_2d}$, $E_N^{a_2n}$, $E_N^{de}$, and $E_N^{ne}$ subsystems lie around the detuning region when cavity 1 is resonant with the anti-Stokes sideband while cavity 2 is resonant with the  Stokes sideband.
However, the peaks of $E_N$ curves representing their entanglement gradually shift from $\Delta_a\approx-\omega_d$ towards $\Delta_a\approx0$ as we move from $J=0.4\,\omega_d$ to $J=1.4\,\omega_d$ and the region for the existence of entanglement also broadens.
Since we have considered $\Delta_e=-\omega_d$ in Fig.~\ref{fig:7}, $E_N^{a_1d}$ and $E_N^{a_1n}$ entanglement is quite weak in this parametric domain.
Nonetheless, it is apparent that $E_N^{a_1d}$ and $E_N^{a_1n}$ entanglement also increases with increasing $J$ reaching a peak value followed by a decreasing trend [See Appendix~\ref{A} for details].

\begin{table}[b]
\begin{tabular}{|l|c|c|c|c|c|}
\hline
\textbf{Subsystem} & {$\mathbf{\Delta_{1}}$}& \textbf{$\mathbf{\Delta_{2}}$}& \textbf{$\mathbf{\tilde{\Delta}_{n}}$}& \textbf{$\mathbf{\Delta_{e}}$}& \textbf{$\mathbf{J}$}\\
\hline
 $E_N^{a_1n}$&$-1.41\,\omega_d$&$-0.68\,\omega_d$&$0.65\,\omega_d$&$-1.63\,\omega_d$&$0.35\,\omega_d$\\
\hline
 $E_N^{a_1d}$&$-0.04\,\omega_d$&$0.85\,\omega_d$&$0.77\,\omega_d$&$0.99\,\omega_d$&$1.28\,\omega_d$\\
\hline
 $E_N^{ne}$&$0.76\,\omega_d$&$-0.52\,\omega_d$&$0.77\,\omega_d$&$-0.63\,\omega_d$&$0.8\,\omega_d$\\
\hline
 $E_N^{de}$&$0.28\,\omega_d$&$-0.84\,\omega_d$&$0.6\,\omega_d$&$-1.07\,\omega_d$&$1.06\,\omega_d$\\
\hline
\end{tabular}
\caption{Optimized parameters for $E_N^{a_1n}$, $E_N^{a_1d}$, $E_N^{ne}$, and $E_N^{de}$ used in Fig.~\ref{fig:9}.}
\label{table:2}
\end{table}

\begin{figure}[h!]
	\includegraphics[width=\linewidth]{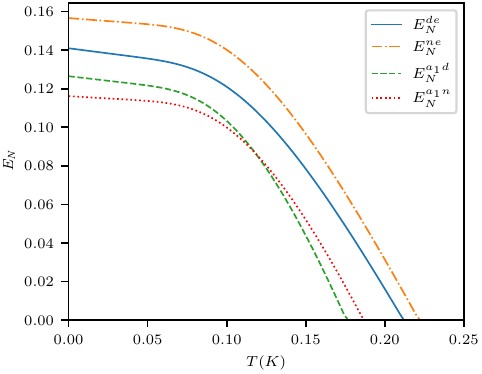}
	\caption{Line plots for $E_N^{de}$, $E_N^{ne}$, $E_N^{a_1d}$, and $E_N^{a_1n}$ as a function of temperature each considered at the optimized value of cavity detuning $\Delta_1$ and $\Delta_2$, effective magnon detuning $\tilde{\Delta}_n$, ensemble detuning $\Delta_e$, and cavity-cavity coupling rate $J$ as shown in Table~\ref{table:2}.}\label{fig:9}
\end{figure}

Next, we present the results of our numerical simulations demonstrating the critical temperature ($T_c$) for $E_N^{de}$, $E_N^{ne}$, $E_N^{a_1d}$, and $E_N^{a_1n}$ in Fig.~\ref{fig:9}.
Entangled subsystems magnon-ensemble and phonon-ensemble exhibit the most robust entanglement against temperature, which can last up to $200\mathrm{~mK}$.
On the other hand, cavity 1 photon-magnon entanglement can survive temperatures up to $180\mathrm{~mK}$.
However, cavity 1 photon-phonon subsystem can sustain their entanglement at as high a temperature as $170\mathrm{~mK}$.
Each curve in Fig.~\ref{fig:9} is plotted for an optimized set of parameter values given in Table~\ref{table:2}.
\begin{widetext}
\onecolumngrid
 \begin{figure}[htb] 
\includegraphics[width=1\linewidth]{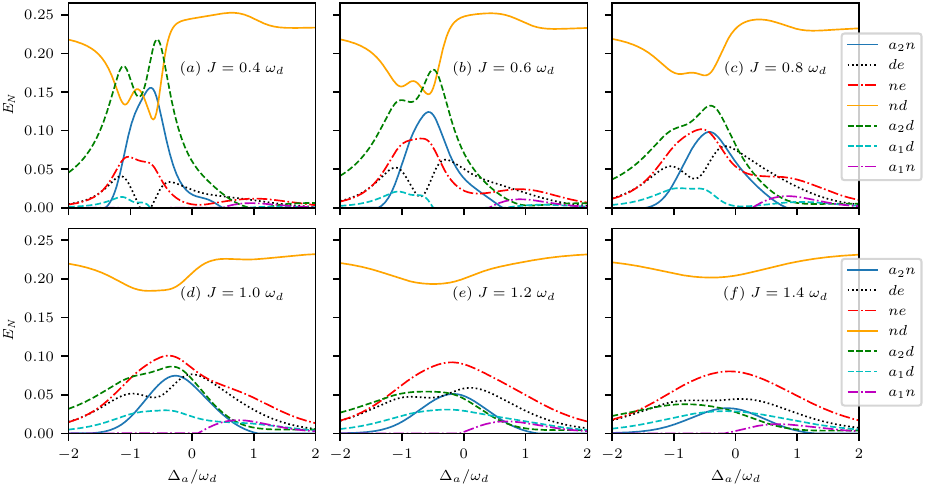}
\caption{Line plot illustrating the effect of cavity-cavity coupling rate $J$ against cavity detuning $\Delta_a/{\omega_d}=-\Delta_1/{\omega_d}=\Delta_2/{\omega_d}$ on bipartite entanglement of cavity 2 photon-magnon ($a_2n$), phonon-ensemble ($de$), magnon-ensemble ($ne$), magnon-phonon ($nd)$, cavity 2 photon-phonon ($a_2d$), cavity 1 photon-phonon ($a_1d$), and cavity 1 photon-magnon ($a_1n$) varied in regular intervals from $J=0.4\,\omega_d$ to $J=1.4\,\omega_d$ in (a)-(f) at $\tilde{\Delta}_n=0.9\,\omega_d$ and $\Delta_e=-\omega_d$.}\label{fig:7}
\end{figure}
\end{widetext}

\twocolumngrid

\begin{figure}[h!]
	\includegraphics[width=\linewidth]{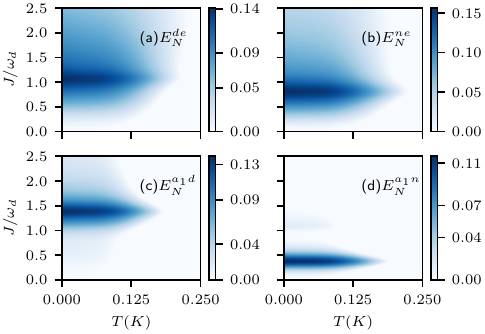}
	\caption{Density plot of (a) $E_N^{de}$, (b) $E_N^{ne}$ (c) $E_N^{a_1d}$, and (d) $E_N^{a_1n}$ as a function of temperature $T$ and normalized coupling rate $J/\omega_d$. Other parameters are optimized as shown in Table~\ref{table:2}.}\label{fig:10}
\end{figure}

It is important to find how the strength of cavity-cavity coupling $J$ impacts the robustness of distant entanglement against temperature.
In Fig.~\ref{fig:10}, we present density plots of $E_N^{de}$, $E_N^{ne}$, $E_N^{a_1d}$, and $E_N^{a_1n}$ as a function of temperature $T$ and cavity-cavity coupling $J$.
We infer from Fig.~\ref{fig:10} that $T_c$ for the existence of entanglement varies with $J$. The maximum value of $J$ corresponding to maximal $E_N^{de}$ [see Fig.~\ref{fig:10}(a)], $E_N^{ne}$ [see Fig.~\ref{fig:10}(b)], $E_N^{a_1d}$ [see Fig.~\ref{fig:10}(c)], and $E_N^{a_1n}$ [see Fig.~\ref{fig:10}(d)] is $1.06\,\omega_d$, $0.8\,\omega_d$, $1.28\,\omega_d$, and $0.35\,\omega_d$, respectively. We observe that $T_c$ is maximum, corresponding to $J$ for which the degree of entanglement is maximal at $T=0$. 
Hence, we can say $T_c$ can be increased through a proper choice of parameters. 

\begin{figure}[htb] 
	\includegraphics[width=\linewidth]{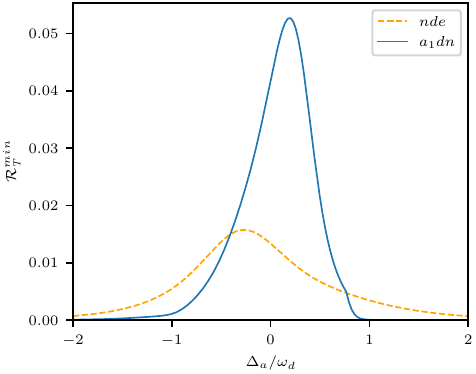}
	\caption{Tripartite entanglement of cavity 1 photon-magnon-phonon ($a_1dn$) and ensemble-magnon-phonon ($nde$) modes as a function of $\Delta_a/{\omega_d}=-\Delta_1/{\omega_d}=\Delta_2/{\omega_d}$ at $\Delta_e=2\,\omega_d$, $\tilde{\Delta}_n=0.25 \,\omega_d$ for $a_1dn$ and $\Delta_e=-0.5\,\omega_d$, $\tilde{\Delta}_n=0.65\,\omega_d$ for $nde$ tripartite subsystems. Cavity-cavity coupling strength is $J=\omega_d$.}\label{fig:8}
\end{figure}
Apart from the bipartite entanglement of different subsystems in coupled magnomechanical system, we show that genuine tripartite entanglement can also be realized for indirectly coupled subsystems.
The same magnomechanical system without cavity 1 was recently considered by Jie Li \textit{et al.}~\cite{li_magnon-photon-phonon_2018} in which they showed that the tripartite magnon-phonon-photon entanglement exists when $\tilde{\Delta}_n\simeq0.9\,\omega_d$ (anti-Stokes sideband) and $\Delta_a\simeq -\omega_d$ (Stokes sideband).
In the coupled magnomechanical system, we consider magnon-phonon-ensemble ($nde$) and cavity-1 photon-phonon-magnon ($a_1dn$) tripartite subsystems and plot the minimum of the residual contangle in Fig.~\ref{fig:8} as a function of normalized detuning $\Delta_a/\omega_d$.
Both these entanglements $nde$ and $a_1dn$ exist for a significant range of cavity field detuning with maximum values near the resonant frequency.
Interestingly, cavity-1 photon-phonon-magnon entanglement has approximately the same degree of entanglement as found in single cavity case~\cite{li_magnon-photon-phonon_2018}.
\\
In the coupled cavity scheme, future investigations may incorporate the inclusion of cross-Kerr non-linearity~\cite{ShenPRL2022, sheng_efficient_2008}, exploration of entanglement dynamics in ultra-strong coupling regime~\cite{teo_double_2013}, study of Einstein-Podolsky-Rosen (EPR) steering~\cite{tan_einstein-podolsky-rosen_2021}, and the introduction of an optical parametric amplifier(OPA) to widen the parametric regime for entanglement~\cite{hussain_entanglement_2022}.
Furthermore, the noise-induced decoherence can be curtailed by purification and entanglement concentration in a practical long-distance quantum communication network~\cite{bennett_purification_1996, yamamoto_concentration_2001}. 

\section{Conclusion}
We proposed a scheme to realize distant entanglement between various excitation modes of the YIG sphere, atomic ensemble, and microwave modes of two coupled cavities housing an atomic ensemble and a YIG sphere.
We have shown that ensemble-phonon and ensemble-magnon distant bipartite entanglements not only exist but can sustain up to $200\,$mK temperature, for a proper choice of experimentally feasible parameters.
Similarly, the entanglement of magnon and phonon modes with cavity-1 photons is also robust against a temperature of about $170\,$mK.
Most importantly, we demonstrate that two types of tripartite entanglement between different distant modes are possible in the proposed system.
These include magnon-phonon-ensemble and cavity-1 photon-phonon-magnon entanglements.
Interestingly, the strength of cavity-1 photon-phonon-magnon tripartite entanglement is comparable to the originally proposed photon-phonon-magnon tripartite entanglement of the same cavity modes.
Hence, we conclude that both the bipartite and tripartite entanglements between indirectly coupled systems are found to be substantial in our proposed setup.
Moreover, cavity-cavity coupling strength plays a key role in the degree of entanglement as well as the range of parameters in which it subsists.
We believe that the parametric regimes identified in our proposed system may prove useful for the experimental realization of distant entanglement, which is significant for processing continuous variable quantum information in quantum memory protocols.

\appendix
\section{Entanglement Transfer}
\label{A}
\begin{figure}[h]
	\includegraphics[width=0.9\linewidth]{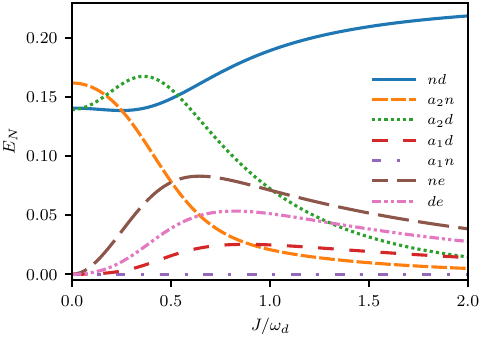}
	\caption{Line plot of $E_N$ against cavity-cavity coupling rate $J$. The rest of the conditions and parameters are the same as in Fig.~\ref{fig:7} with $\Delta_a=-\omega_d$.}\label{fig:enj}
\end{figure}

\begin{figure}[h]
	\includegraphics[width=0.9\linewidth]{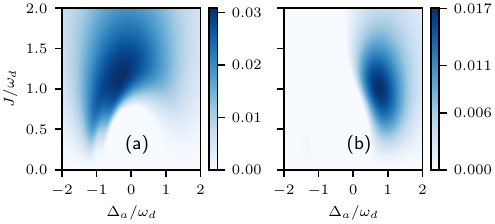}
	\caption{Density plot illustrating the effect of cavity-cavity coupling rate $J$ and cavity detuning $\Delta_a$ on bipartite entanglement of (a) cavity 1 photon-phonon ($a_1d$) and (b) cavity 1 photon-magnon ($a_1n$). The rest of the conditions and parameters are the same as in Fig.~\ref{fig:7}.}\label{fig:ent1}
\end{figure}

Here, we further discuss the entanglement transfer phenomenon previously studied in Fig.~\ref{fig:7}.
In Fig.~\ref{fig:enj}, we plot bipartite entanglements as a function of cavity-cavity coupling $J$ at $\Delta_a=-\omega_d$.
It can be seen, that when $J=0$, we only have entanglement between three modes of cavity-2 since cavity-1 is decoupled.
When cavity-cavity coupling is turned on, the cavity fields interact with each other.
As a result the population of various modes changes leading to the transfer of entanglement between different modes.
Fig.~\ref{fig:enj} shows that initially $E_N^{nd}$ and $E_N^{a_2n}$ decreases with increase in $E_N^{a_2d}$, $E_N^{a_1d}$, $E_N^{ne}$, and $E_N^{de}$.
A further increase in $J$ leads to a decreasing trend in $E_N^{a_2d}$.
On the other hand cavity 1 photon-magnon ($E_N^{a_1n}$) entanglement increases as a function of $J$ around $\Delta_a\approx\omega_d$, reaching a maximum followed by a decaying trend as shown in Fig.~\ref{fig:ent1}(b).
Similarly, Fig.~\ref{fig:ent1} (a) illustrates a similar trend for the cavity 1 photon-phonon, reaching a peak value near resonance followed by the decaying trend.
Besides the increase in entanglement amplitude, the domain of entanglement in detuning space also increases.

\bibliography{References}
\end{document}